\documentstyle[preprint,aps]{revtex}

\begin{document}
  \draft
\preprint{\small gr-qc/9404047 Alberta-Thy-12-94}

 \title
 {Radiation collapse and gravitational waves\\
 in three dimensions}
\author{Viqar Husain}
\address
 { Theoretical Physics Institute,\\
 University of  Alberta,
Edmonton, Alberta, Canada T6G 2J1}
\maketitle

\begin{abstract}
Two non-static solutions for three dimensional gravity coupled to
matter fields are given.  One  describes  the collapse of radiation
that results in a black hole. This is the three dimensional analog
of the Vaidya metric, and is used to construct a model for
mass inflation. The other describes plane gravitational waves
for coupling to a massless scalar field.
 \end{abstract}

\vfill
\eject

Lower dimensional gravity  has often been used as an arena
for  investigating various problems that arise in four dimensions,
but  are not solvable there.  Among those that have been
substantially
investigated    include quantum gravity in three  dimensions
\cite{ash,wit} and
black hole evaporation in  two dimensions \cite{donrev}.

Obtaining classical solutions  in lower dimensions
is often a first step in these models.  In three dimensional gravity
the
solutions
for point masses were the first to be studied \cite{des}.   More
recently a
   black hole solution has been given  by Banados et. al.
\cite{btz},   which also provides an arena for investigating black
hole
evaporation.

There are two classical problems in general relativity in four
dimensions that
have  recently attracted  some  attention, and a  three dimensional
version of

them may be useful to address.

The first  problem has to do with the   inner (Cauchy) horizon of
the
Kerr  and Reissner-Nordstrom   black holes.   This horizon is
believed to be
 unstable to time dependent perturbations because it is a surface
where
infalling radiation is infinitely  blue shifted.
The question is what effect the back reaction of the blue shifted
radiation
has on the internal geometry.  More precisely one would like to know
what type of singularity develops at or before the Cauchy horizon as
a result
of this back reaction.
This question is important for the cosmic censorship hypothesis, for
if the Cauchy horizon can be crossed, the timelike curvature
singularity in
such spacetimes
becomes naked.  This question may   be asked of  plane wave
spacetimes,
which  also have Cauchy horizons.

Recent approaches to this problem, within spherical symmetry,  take
into
account  non-linear perturbations at the Cauchy horizon as well as
their back
reaction on the geometry.   The results suggest that  the singularity
has a
null
portion, where the internal mass function of the black hole
diverges \cite{pi,ori}.   However it is not yet known  what  type of
singularity  replaces the  Cauchy horizon under general
perturbations.

The second problem is the   investigation of  the collapse of
matter fields to form black holes.   This has been studied
numerically
 and the results are intriguing \cite{chop,ae}.   It has been found
that
when the initial matter field  is an ingoing pulse, the collapsing
matter forms
a black  hole with  mass  given by $M = K(c-c_*)^\gamma$, where  $K$
is
a constant,  $c$ is any one of the parameters in the initial data
for the matter field,  $c_*$ is the critical value of this parameter
(that gives a zero mass black hole), and $\gamma\sim .36$.  In
particular,   no black hole is formed when $c<c_*$.   An
important feature of this result is that it appears to be independent
 of spherical symmetry and the type of matter fields,  with the same
 numerical  exponent $\gamma$ appearing in all cases studied to date.

 This  seems to reflect a universal property of the Einstein
equations in

strong
 field regions.   So far there is no analytical understanding of this
result.
  It would be  interesting  to see if  a similar result   is true  in
lower
dimensions,  and whether  it can be better understood
there, perhaps analytically.

This paper is concerned   with the first problem,  and two metrics
 in three dimensions are given that have Cauchy horizons.   One has
the Vaidya form and describes  collapsing spherically symmetric
radiation.
It allows the construction of  the Ori model  for mass inflation
\cite{ori} in
three dimensions .  The  other describes plane gravitational waves
for
coupling to a massless scalar field.   This metric may also be used
as a
starting point for studying  perturbations of Cauchy horizons
\cite{yurt}.

The Vaidya form of a  three dimensional  metric may be written
using an advanced time  coordinate $v$, and polar coordinates
$r,\theta$
in the plane.  It is
\begin{equation}
ds^2 = -f(r,v)dv^2 + 2drdv + r^2 d\theta^2  \label{met}.
\end{equation}
The total energy momentum tensor we use contains contributions
\begin{equation}
I_{\alpha\beta}  = {\rho(v) \over 4\pi r} \partial_\alpha v
\partial_\beta v
\end{equation}
for   infalling   radiation  with luminosity $\rho(v)$,  and
\begin{equation}
E_\alpha^\beta  =   {q^2\over 4\pi r^2} {\rm diag}(-1,-1,1)
\end{equation}
(in the coordinates $(v,r,\theta)$) for the external electric field
due to
a charge $q$.

 The Einstein equations with cosmological constant $\Lambda$
\begin{equation}
G_{\alpha\beta } + \Lambda g_{\alpha\beta }=2\pi( I_{\alpha\beta } +
E_{\alpha\beta } )
\end{equation}
 have,  with the ansatz (\ref{met}),   the solution
\begin{equation}
f(r,v)  = - [\Lambda r^2 + g(v) +  q^2 {\rm ln} r ], \label{sol1}
\end{equation}
where  $g(v)$ is given by
\begin{equation}
{dg(v)\over dv} = \rho(v). \label{sol2}
\end{equation}
This gives the three dimensional  analog of the charged Vaidya metric
\cite{vaid}.

If asymptotically ($r\rightarrow \infty$ and $v\rightarrow -\infty$)
the
radiation
inflow vanishes so that $g(v)=0$,  the metric assumes the  `vacuum'
form
\begin{equation}
ds^2 =  (\Lambda r^2  + q^2 {\rm ln}r) dv^2 + 2 dvdr  + r^2
d\theta^2.
\label{vac}
\end{equation}
This shows that  $\Lambda\equiv -l^{-2}$  must be negative for the
metric to be
asymptotically Lorentzian.

  The case
$\rho(v)  = 0$,   $g(v) = M$,  a constant,  gives the spherically
symmetric
static
black hole found by Banados et. al \cite{btz}. The event horizon (for
$q=0$)
is  at
\begin{equation}
r_{EH}= l \sqrt {M}.
\end{equation}
  When $\rho(v)\ne 0$,  the mass of the black hole  formed from the
collapse
  depends on the parameters in the ingoing pulse.  Asymptotically
 ($v\rightarrow \infty$), the apparent horizon becomes null and its
radius gives the black hole mass.
The radial coordinate of the apparent horizon $r_{AH}$ is a measure
of the
 black hole mass function $m(v)$, and is given  (again for $q=0$) by
\begin{equation}
m(v):= r_{AH}(v) = l\sqrt{g(v)}.
\end{equation}
 As an  example,  a   `soliton'  form for the radiation inflow
\begin{equation}
\rho(v) = A{\rm sech}^2 {v\over b}  \label{solit}
\end{equation}
 gives
\begin{equation}
g(v) =  Ab{\rm tanh}{v\over b},
\end{equation}
where $A,b$ are constants.  The radius of the apparent horizon  in
the
$v\rightarrow \infty$ limit gives  the black hole mass formed in the
collapse.
The inflow   (\ref{solit}) gives
\begin{equation}
m_{BH} := \lim_{v\rightarrow \infty}  r_{AH} = l \sqrt{Ab}.
\end{equation}

  The charged case  is  interesting   when $g(v)=0$, since then
 \begin{equation}
f(r) = (r/l)^2 - q^2 {\rm ln}r.
\end{equation}
  There is now the possibility of  horizons
 even for zero mass depending on the value of  the
product $ql$.  This is unlike the case for four dimensional
Reissner-Nordstrom
black holes  with  negative cosmological constant  $\Lambda = -l^2$
 where, for  zero mass,
\begin{equation}
f = 1+ ({ q \over r})^2 + {1\over 3} ({r \over l})^2,
\end{equation}
 and the singularity at  $r=0$  is naked.
In the present case however, the equation for the horizon is
\begin{equation}
({r\over l })^2 = q^2  {\rm ln}r,
\end{equation}
which first has a single solution  for  $ql= \sqrt{2e}$.  This
corresponds to
a
charged massless black hole without an inner horizon. For
$ql < \sqrt{2e}$ the $r=0$ singularity is naked, whereas
for   $ql > \sqrt{2e}$ there is both an inner and outer horizon.

 As mentioned above, there have been analytical studies of the nature
of the
Cauchy horizon designed to take into account  the back reaction  of
blue shifted radiation on the internal black hole geometry.    It has
been
found that  the internal  mass function diverges at the Cauchy
horizon  - a
phenomenon which has been dubbed  `mass inflation' \cite{pi}.

 The inputs that yield this result   for charged  spherically
symmetric black
holes are (i) infalling radiation, and (ii) `backscattered'
outgoing
radiation, all
within the event  horizon. The effect of the  outgoing  radiation  is
that it
displaces
 the Cauchy  horizon on the outer side of the outgoing pulse
relative to that
on
its inner side.  This is crucial to obtaining  mass inflation.

These ingredients are  most easily realized in a model due to Ori
\cite{ori}
constructed by patching  together   two Vaidya metrics  along an
outgoing null
surface, with each metric describing infalling radiation.  The
outgoing
radiation
is thus modelled as the null shell $\Sigma$ where  the  metrics are
glued
(see the figure).

This model may be constructed  in three dimensions using the metric
(\ref{met}).  The main steps are identical to the four dimensional
case
\cite{ori} .  The two relevant equations  are,  (i) the equation
of the outgoing null shell obtained from (\ref{met})
\begin{equation}
   f_+dv_+ = f_-dv_- = 2dr, \label{shell}
\end{equation}
where the subscripts $\pm$ denote the inner (outer) Vaidya metrics,
and
(ii) the continuity of the inflow  along the null shell.  This last
condition
is obtained by requiring  the continuity of  the components of the
energy-momentum  tensor along the null rays
\begin{equation}
n_\pm^\alpha = ({2\over f_\pm},1,0):
\end{equation}
\begin{equation}
T_{\alpha\beta} n^\alpha n^\beta |_+ =
T_{\alpha\beta} n^\alpha n^\beta |_- ,
\end{equation}
which gives
\begin{equation}
 {1\over f_+^2} {dg_+\over dv_+} = {1\over f_-^2} {dg_-\over dv_-}.
\label{em}
\end{equation}
Combining (\ref{shell}) and (\ref{em}) gives the key relation that
gives
rise to mass inflation:
\begin{equation}
{dg_+\over f_+} = {dg_-\over f_-}.
\end{equation}
At the Cauchy horizon on the outer side of the outgoing null shell
($v_-\rightarrow \infty$,   $f_-\rightarrow 0$),  the right hand
side diverges since $g_-$, which is determined by the inflow, is
assumed
to be  bounded.  Also
$f_+$ is bounded because of the presence of the outgoing shell, which
shifts  the vanishing of  $f_+$ to a smaller value of the radial
coordinate
(see the figure).
Thus $g_+$, which gives  the internal mass must  diverge.  As done in
the
 four  dimensional case,  $g_+$ may be computed
explicitly  by assuming a specific power law  fall off for $\rho_-$,
and using
the above equations.  This shows that   mass inflation also occurs in
three
dimensions.  Similar results have been obtained for two dimensional
dilatonic black holes \cite{serge}.

Another metric in three dimensions that contains Cauchy horizons  is
one that describes gravitational waves when there is coupling to
matter fields.  (Without matter fields, three dimensional gravity has
only
 a finite number of degrees of freedom so there can be  no
gravitational
waves).

Here we consider coupling to a massless scalar field and give a
gravitational  wave solution.  The  Einstein equation for  scalar
field
coupling  may be written as
\begin{equation}
R_{\mu\nu} = 2\pi  \phi_\mu \phi_\nu.
\end{equation}
With the ansatz
\begin{equation}
ds^2 = -dt^2 + dx^2 +  a^2(x,t) dy^2,
\end{equation}
the field equations become
\begin{equation}
\ddot{a}  - a'' = 0,
\end{equation}
\begin{equation}
{a''\over a }  = -2\pi (\phi')^2, \ \ \ \ \ \
{\ddot{a} \over a } = -2\pi \dot{\phi}^2, \ \ \ \ \ \
{\dot{a}'\over a} = -2\pi \dot{\phi}\phi',
\end{equation}
where the dot and prime denote $t$ and $x$ derivatives.
The solutions to these equations are ingoing or outgoing
waves $a(v), a(u)$  where  $v=t+x$ and $u=t-x$.   The scalar
field  $\phi$ and $a$ are connected by the latter three equations,
all of
which  reduce to one equation.   An explicit example is the scalar
field
linear in $v$ which gives sinusoidal waves.   The Cauchy horizon is
the null
surface   where $a(v)=0$.

 When  considering  general non-linear perturbations
of  the Cauchy horizon,  arguments similar to those given in ref.
\cite{yurt}
for
 four dimensional  plane wave spacetimes apply
in this case as well.  But,  unlike the black hole case,  there is as
yet no
simple model for studying  perturbations  of the Cauchy horizon for
these
spacetimes.

In summary, the first metric  given here generalizes the three
dimensional
black hole \cite{btz} to the non-static case, and  gives a model for
mass
inflation.  The second metric  gives gravitational  wave solutions
when there
is  coupling to a massless scalar field.   As discussed above, these
metrics
may be of use in studying  the effect of general perturbations on the
Cauchy horizon in a simpler setting.

I would like to thank   Warren Anderson, Serge Droz, Werner Israel,
and
 Sharon Morsink  for  comments and  discussions.

\end{document}